\documentclass[11pt,a4paper]{article}

\usepackage[margin=2.5cm]{geometry}
\usepackage[utf8]{inputenc}
\usepackage[T1]{fontenc}
\usepackage{amsmath,amssymb}
\usepackage{graphicx}
\usepackage{array}
\usepackage{tabularx}
\usepackage{longtable}
\usepackage{rotating}
\usepackage{pdflscape}
\usepackage[numbers,sort&compress,merge]{natbib}
\bibpunct[, ]{(}{)}{,}{n}{,}{,}
\usepackage{orcidlink}
\usepackage{hyperref}
\hypersetup{colorlinks=true, linkcolor=blue, citecolor=blue, urlcolor=blue}

\usepackage{newunicodechar}
\newunicodechar{−}{\ensuremath{-}}
\newunicodechar{α}{\ensuremath{\alpha}}
\newunicodechar{β}{\ensuremath{\beta}}
\newunicodechar{ε}{\ensuremath{\varepsilon}}
\newunicodechar{₁}{\ensuremath{_1}}
\newunicodechar{₂}{\ensuremath{_2}}
\newunicodechar{₃}{\ensuremath{_3}}
\newunicodechar{₄}{\ensuremath{_4}}
\newunicodechar{₅}{\ensuremath{_5}}
\newunicodechar{ᵢ}{\ensuremath{_i}}

\title{Inside the Degree, Outside the Discipline? Testing an Asymmetric Appraisal Model of the Curricular Legitimacy Gap in Computing Education}

\author{
Md Abdullah Al Kafi\textsuperscript{1,2}\orcidlink{0009-0006-4361-029X},\quad
Walayat Hussain\textsuperscript{2,*}\orcidlink{0000-0003-0610-4006},\quad
Raka Moni\textsuperscript{1}\orcidlink{0009-0008-5391-9893}
\\[0.6em]
\small\textsuperscript{1}Multidisciplinary Action Research (MARS) Lab, Daffodil International University,\\
\small Savar, Dhaka 1216, Bangladesh\\[0.2em]
\small\textsuperscript{2}Artificial Intelligence for Decision Excellence (AIDX) Lab, Peter Faber Business School,\\
\small Australian Catholic University, 40 Edward Street, North Sydney, NSW 2060, Australia\\[0.4em]
\small\textsuperscript{*}Corresponding author: \texttt{Walayat.Hussain@acu.edu.au}
}
\date{}

\begin{document}

\maketitle

\begin{abstract}
\noindent
Required broader coursework can secure participation without being recognised as legitimate computing knowledge. This study conceptualises this disconnect as a curricular legitimacy gap and tests an asymmetric appraisal model grounded in situated expectancy value theory. The model distinguishes curricular devaluation, judging broader coursework unnecessary or professionally irrelevant, from integrative intention, or willingness to reuse its learning. Survey data from 212 Computer Science and Engineering undergraduates in Bangladesh recruited through snowball sampling were analysed using robust structural equation modelling. Primary inference combined robust direct-path estimates with 5,000 respondent-level bootstrap resamples; alternative measurement, response-quality, and ordinal-estimator specifications were also examined. Perceived burden was positively associated with devaluation, which was negatively associated with intention. The standardised indirect association of burden with intention through devaluation was −.350, 95\% CI [−.589, −.169]. Perceived benefits were associated with stronger intention through lower devaluation, indirect association .141, 95\% CI [.058, .253], and an additional positive direct pathway. The model explained 47.6\% of the variance in devaluation and 40.6\% in intention. The direct burden-to-intention pathway was unsupported under the primary estimator but significant in the opposite-to-hypothesised direction under ordinal estimation; this residual path is therefore treated as estimator-dependent. The results support a distinction among requirements, valuation, intention, and behaviour and suggest that cost reduction and utility development address different curricular problems. Given the cross-sectional, nonprobability design and developing measures, all pathways are interpreted as associations rather than causal mediation.

\vspace{0.5em}
\end{abstract}

\section{Introduction}
Engineering problems rarely arise as purely technical tasks. Computing systems operate within institutions, influence diverse communities, and produce consequences that extend well beyond functional performance. Decisions concerning system design, data use, automation, accessibility, privacy, and artificial intelligence require technical competence alongside communication, ethical judgement, social awareness, and an understanding of organisational and human contexts. Reflecting these professional demands, the Computer Science Curricula 2023 guidelines position society, ethics, and professional responsibility as integral elements of computing education rather than optional additions to technical preparation \cite{Kumar2024}. The central educational challenge is therefore not simply whether computing programmes should include broader forms of learning. It is whether students recognise such learning as a legitimate and professionally consequential part of becoming a computing professional.
A common curricular response has been to require courses in communication, ethics, humanities, social sciences, management, and related areas. These requirements can expand the intellectual and professional scope of computing education. Evidence from interdisciplinary educational interventions suggests that integrating humanities and engineering perspectives can strengthen students’ attention to people, broader contexts, and the interaction between social and technical considerations \cite{Czerwionka2025}. Research on interdisciplinary education similarly shows that meaningful integration requires students to connect different forms of knowledge, assumptions, and methods rather than merely encounter content from several disciplines \cite{Ming2026, Spelt2009}. Curricular inclusion is therefore necessary, but it is not sufficient. A programme can require broader coursework without persuading students that this knowledge belongs within computing practice.
This distinction is important because formal participation does not necessarily produce intellectual commitment. Students may complete a required course while regarding it as peripheral, unnecessary, or less valuable than programming, mathematics, and systems subjects. Such judgements can influence the effort students invest in the course and their willingness to apply its learning in later projects, internships, or professional decisions. Studies of engineering culture have shown that students may encounter a hierarchy of knowledge in which technical expertise is treated as central while ethical, social, and public welfare considerations are positioned as secondary \cite{Cech2014, Lim2021, Niles2020}. Even programmes that explicitly promote public welfare can face difficulty convincing students that social and ethical reasoning constitutes legitimate engineering work \cite{Niles2020}.
This study defines the resulting disconnect as the curricular legitimacy gap. The term refers to the distance between an institution’s decision to require broader learning and a student’s judgement that such learning properly belongs within professional computing education. This gap is more specific than dissatisfaction with a course and more conceptually precise than general resistance. It concerns the status that students assign to a curricular domain. A student may comply with a requirement without valuing it, value it without intending to use it, or intend to use it without subsequently transferring the learning into practice. Requirement, valuation, intention, and behaviour must therefore be treated as related but distinct educational states.

\subsection{Research Gap}
Existing scholarship provides important insights into this problem, but the relevant explanations remain fragmented. Curriculum research establishes the professional need for social, ethical, communicative, and interdisciplinary learning. Intervention studies examine whether particular educational designs develop broader forms of thinking. Research on engineering culture demonstrates how disciplinary norms can marginalise social and ethical concerns. Motivation research explains how students evaluate the benefits and costs of academic tasks. However, these streams do not yet offer a sufficiently integrated explanation of how students convert perceived academic costs and benefits into a judgement about whether a curricular domain is legitimate, or how that judgement is associated with their willingness to integrate the learning into future practice.
Recent computing education reviews sharpen this unresolved problem. \cite{brown2024teaching} mapped 100 publications on ethics teaching across ACM computing education venues and found substantial variation in content, pedagogy, and evaluation, with evaluations relying heavily on student self-reports, course evaluations, and instructor reflections. Padia et al realist review further showed that students’ acceptance of ethical interventions in technical computing courses is context-dependent and identified four explanatory routes across 13 reports \cite{Padiyath2024}. Together, these reviews establish that student response to sociotechnical instruction is neither automatic nor unitary. Their primary unit of explanation, however, is the ethics intervention. They do not distinguish an institution’s requirement from students’ valuation of a broader curricular domain, their intention to reuse its learning, and later behavioural integration, or test whether cost and utility appraisals reach intention through different pathway configurations. That is the specific gap addressed in this study.
This omission creates an important conceptual problem. If participation is treated as acceptance, compulsory enrolment can be mistaken for educational uptake. If resistance is attributed primarily to workload, programmes may respond by reducing assessment demands without addressing students’ doubts about professional relevance. Conversely, if positive attitudes are treated as evidence of transfer, the distinction between recognising value and intending to apply learning becomes obscured. What remains insufficiently examined is the evaluative process connecting students’ experience of a course with their willingness to carry its learning forward.
Situated expectancy value theory provides a useful lens for explaining this process. The theory proposes that students’ choices and engagement are shaped by their expectations, the value they assign to an activity, and the costs they associate with participation. These evaluations are situated within social messages, prior experiences, identities, and future goals \cite{eccles2020expectancy}. Research on academic cost further demonstrates that perceived cost is multidimensional and conceptually distinct from perceived value \cite{Flake2015}. A course can therefore be demanding but professionally valuable, or relatively manageable but perceived as irrelevant. Cost and value should not be treated as opposite ends of a single continuum.
Applied to broader computing coursework, this distinction suggests an asymmetric appraisal process. Perceived burden may not reduce integrative intention simply because the course requires effort. Its influence may become consequential when students interpret that effort as being spent on learning that does not belong within their professional formation. Curricular devaluation may therefore provide the principal pathway through which burden is associated with weaker integrative intention. Perceived benefits may operate differently. By making professional utility visible, benefits may reduce devaluation while also supporting integrative intention more directly. This model offers a more precise explanation than the assumption that burden and benefit exert equal and opposite effects.
The present analysis is informed by situated expectancy value theory but does not claim to test the complete theory. The study does not measure expectancy of success or each established dimension of task value. Perceived academic burden is treated as an overall appraisal of academic cost, while perceived benefits represent the utility that students associate with broader coursework. The theoretical framework is therefore used to specify the relationship among cost appraisal, benefit appraisal, curricular valuation, and intended integration.
Construct precision is equally important. The items originally associated with academic resistance primarily assess whether broader courses are perceived as unnecessary, irrelevant, or a poor use of time. They do not directly measure protest, avoidance, emotional opposition, or resistant behaviour. The construct is therefore described as curricular devaluation. Similarly, the outcome items assess openness to integration, perceived applicability, and intended future use. They do not measure observed engagement or the actual transfer of learning into professional practice. The outcome is consequently described as integrative intention. These distinctions align the theoretical claims with what the survey can reasonably support.
This study examines these relationships using survey data from 212 Computer Science and Engineering undergraduates in Bangladesh. The context extends an evidence base that has largely been developed through studies of engineering education in North American and European settings. It also provides a focused setting in which to examine the perceived legitimacy of broader coursework within a technically intensive degree programme. The questionnaire uses the category “non-STEM” to describe a heterogeneous group of courses. This term is retained only when referring to the wording presented to participants. In the conceptual discussion, broader coursework and sociotechnical curricular breadth are used because they describe the intended educational function of these courses without defining them solely through exclusion from STEM.

\subsection{Research Question and Contributions}
The study addresses the following research question:
\textbf{RQ1.} How are perceived academic burden and perceived benefits associated with Computer Science and Engineering students’ curricular devaluation and integrative intention, and what indirect associations operate through curricular devaluation?
This study makes three connected contributions. First, it conceptualises the curricular legitimacy gap as a distinction among institutional requirement, student valuation, intended reuse, and demonstrated integration. This sequence prevents compulsory participation from being interpreted as educational acceptance and prevents intention from being presented as observed transfer. Second, it specifies and evaluates an asymmetric appraisal model. Curricular devaluation is positioned as the evaluative link through which burden is associated with intention, while perceived benefits may relate to intention both by weakening devaluation and through an additional positive route. Third, it replaces broad labels of resistance and acceptance with curricular devaluation and integrative intention, terms that correspond more closely to the observed indicators.
The study does not claim to be the first investigation of computing students’ attitudes towards ethics, professional skills, or broader learning. Its novelty lies in reorganising that problem around a different explanandum: whether an institutionally required curricular domain is accorded professional legitimacy and how that valuation is associated with intended reuse. The contribution is therefore the compliance, valuation, intention, and behaviour distinction together with the path-specific comparison of burden and benefit, not the national setting or the use of structural equation modelling alone.

\section{Theoretical Framework and Hypotheses}
This section develops the theoretical basis for the hypotheses examined in the remainder of the study. Section~\ref{2.1} distinguishes curricular presence from curricular legitimacy and introduces the curricular legitimacy gap as the study's conceptual focus. Section~\ref{2.2} draws on situated expectancy-value theory to motivate perceived academic burden and perceived benefits as distinct cost and utility appraisals and states H1 and H2. Section~\ref{2.3} links curricular devaluation to integrative intention and states H3. Section~\ref{2.4} then specifies the asymmetric appraisal model, in which curricular devaluation functions as the principal evaluative pathway linking burden and benefits to intention, and states H4 through H6b.

\subsection{From Curricular Presence to Curricular Legitimacy}
\label{2.1}
Computing curriculum guidance now treats professional, ethical, and social responsibility as part of the knowledge expected of computing graduates \cite{Kumar2024}. Interdisciplinary scholarship makes a related point: educational integration requires learners to connect disciplinary perspectives, not merely encounter them in adjacent courses \cite{Ming2026, Spelt2009}. This distinction separates curricular presence from curricular legitimacy. Presence is established by placing a course in a programme. Legitimacy is established when students judge that its knowledge properly belongs in their formation and can affect consequential technical decisions.
Curricular legitimacy is neither a synonym for course satisfaction nor a claim that all required courses must be easy or enjoyable. It concerns the status assigned to a domain of knowledge. That status is shaped by formal requirements and by the wider instructional environment, including credit weight, assessment, sequencing, faculty language, and whether the learning reappears in technical courses, projects, internships, and capstones. Engineering cultures that detach technical work from public welfare and social analysis can reinforce a hierarchy in which sociotechnical learning is treated as supplementary \cite{Cech2014, Lim2021, Niles2020}. Conversely, authentic humanities and engineering integration can make the relation between people, context, and technical judgement visible \cite{Czerwionka2025}.
The curricular legitimacy gap therefore refers to a mismatch between institutional prescription and student valuation. It is useful because it distinguishes four states that are often conflated: required participation, valuation of the curricular domain, intention to integrate the learning, and demonstrated use in later practice. The present study examines the middle two states. It does not infer acceptance from enrolment and does not infer behavioural transfer from intention. Table~\ref{tab:curricular_legitimacy} presents the definition of four states and the proposed study claim.

\begin{table}[htbp]
\centering
\caption{Conceptual distinctions in the curricular legitimacy framework}
\label{tab:curricular_legitimacy}
\begin{tabular}{p{0.18\textwidth} p{0.42\textwidth} p{0.30\textwidth}}
\hline
\textbf{State} & \textbf{Definition} & \textbf{Measurement status in the present study} \\
\hline
Requirement & Institutional placement of broader coursework within the degree. & Context for the study, not a measured construct. \\
\hline
Curricular valuation & Judgement that the domain is relevant, necessary, and belongs in professional formation. & Measured in its negative form as curricular devaluation. \\
\hline
Integrative intention & Reported openness or intention to apply the learning in later academic, technical, or professional contexts. & Measured as intention, not observed transfer. \\
\hline
Behavioural integration & Demonstrated use of broader learning in decisions, artefacts, projects, or practice. & Not measured and not inferred from the current data. \\
\hline
\end{tabular}
\vspace{0.5em}
\begin{minipage}{0.9\textwidth}
\small
\textit{Note.} The distinction prevents compulsory participation from being treated as educational acceptance and keeps the outcome claim aligned with the survey items.
\end{minipage}
\end{table}

\subsection{Situated Cost and Benefit Appraisals}
\label{2.2}
Situated expectancy-value theory proposes that educational choices reflect expectations, task values, and perceived costs, each situated within identities, prior experiences, social messages, and future goals \cite{eccles2020expectancy}. Cost is not simply low value. Students can regard an activity as demanding and worthwhile, or manageable and irrelevant. Empirical work on academic cost similarly treats effort, lost alternatives, and emotional demands as distinguishable from positive task value \cite{Flake2015}. Students' workload judgements also depend on how they interpret the purpose, organisation, and learning environment surrounding the work, not only on objective study hours \cite{Kember2004}.

In the present model, perceived academic burden summarises cognitive switching, workload pressure, difficulty coordinating technical and broader coursework, and interference with personal time. These experiences should be associated with curricular devaluation when students interpret the required effort as expenditure on knowledge that does not belong in their professional formation. Perceived benefits capture recognised development of communication and critical thinking and a broader perspective on learning. Benefits should weaken devaluation because they make utility visible. These are not mirror image predictions: burden concerns experienced cost, while benefit concerns recognised utility.

\noindent\textbf{H1.} \textit{Perceived academic burden is positively associated with curricular devaluation.}

\noindent\textbf{H2.} \textit{Perceived benefits are negatively associated with curricular devaluation.}

\subsection{Curricular Devaluation and Integrative Intention}
\label{2.3}
Curricular devaluation denotes the judgement that broader coursework is unclear in purpose, difficult to apply, unnecessary, less relevant than technical subjects, or a poor use of time. The construct is evaluative rather than behavioural. Students who assign broader coursework this low status have less reason to retrieve or apply its concepts after the formal requirement ends. Integrative intention, by contrast, captures whether students regard the learning as valuable despite its challenges, remain open to integrating it into academic work, and anticipate applying it in personal or professional life.
The distinction between valuation and intention matters theoretically. A student can acknowledge that a course belongs in the curriculum without committing to use it, just as a student can comply with a course while devaluing it. A negative association is therefore expected between curricular devaluation and integrative intention.

\noindent\textbf{H3.} \textit{Curricular devaluation is negatively associated with integrative intention.}

\subsection{The Asymmetric Appraisal Model}
\label{2.4}
The central theoretical claim concerns the configuration of pathways. If burden matters chiefly because it is interpreted as evidence of curricular misfit, little negative association should remain after devaluation is represented. A direct burden pathway is nevertheless estimated because burden may also reduce willingness to integrate learning through fatigue or competing demands. Perceived benefits may operate through two routes. They can weaken devaluation, and they can provide an affirmative reason for future application even after devaluation is taken into account.

\noindent\textbf{H4.} \textit{Perceived academic burden is negatively associated with integrative intention after curricular devaluation is taken into account.}

\noindent\textbf{H5.} \textit{Perceived benefits are positively associated with integrative intention after curricular devaluation is taken into account.}

\noindent\textbf{H6a.} \textit{Perceived academic burden has a negative indirect association with integrative intention through curricular devaluation.}

\noindent\textbf{H6b.} \textit{Perceived benefits have a positive indirect association with integrative intention through curricular devaluation.}

\begin{figure}[htbp]
\centering
\includegraphics[width=0.5\linewidth]{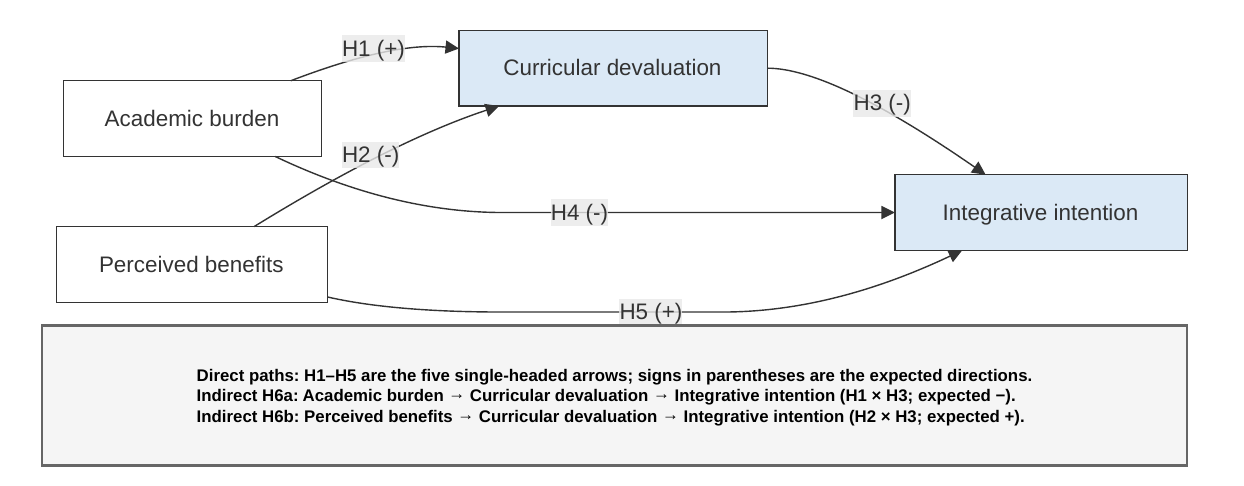}
\caption{Conceptual asymmetric appraisal model}
\label{fig:placeholder}
\end{figure}

\section{Method}
This section describes how the study was designed and conducted and how the data were analysed. Section~\ref{3.1} describes the design, setting, and participants. Section~\ref{3.2} outlines the procedure and the ethical considerations governing data collection. Section~\ref{3.3} details the survey instrument and the operationalisation of each construct. Section~\ref{3.4} reports data preparation and response-quality screening. Section~\ref{3.5} specifies the analytical strategy, including the measurement and structural modelling procedures and the sensitivity analyses used to evaluate the robustness of the findings.
\subsection{Design, Setting, and Participants}
\label{3.1}
Data were collected between 10 April and 10 May 2025 from undergraduate students enrolled in Computer Science and Engineering programmes in Bangladesh. Institutional affiliation was not recorded in the questionnaire or analytical data; the number and identity of universities represented in the respondent sample therefore cannot be established from the available records.
Participants were recruited through non-probability snowball sampling using an online questionnaire administered through Google Forms. The invitation was distributed by the research team through student group chats and personal academic networks, and recipients were encouraged to forward the link to eligible peers. Because the invitation propagated through participants’ own networks, the number of students who received or viewed it could not be determined, and a conventional response rate therefore cannot be calculated. Institutional affiliation was not recorded, so institution-level sample sizes cannot be reported, and institutional clustering could not be examined. Participation was voluntary and recruitment was not probability based, so the sample may be affected by self-selection bias. Students who elected to participate may differ systematically from nonparticipants in their experiences of, or attitudes towards, non-STEM coursework, and snowball propagation may further concentrate the sample within connected peer groups.
The final analytical sample comprised 212 respondents. Of these, 120 identified as men, 48 identified as women, and 44 did not provide a gender response. The sample was concentrated among students in the later stages of their degree: seven respondents were in their second year, 66 were in their third year, and 124 were in their fourth year, while 15 provided an academic-year entry that could not be classified. No first-year students were represented, because students who had not completed at least one full academic year were not eligible to participate. Academic year was collected as an open text response and was subsequently coded to year categories; entries recording a calendar year, a semester beyond the fourth year, or an otherwise uninterpretable value were retained as unclassified rather than assigned to a year. Participants also reported their cumulative grade point average using categorical response options. Table~\ref{tab:participant_characteristics} summarises the demographic and academic characteristics of the sample.

\begin{table}[htbp]
\centering
\caption{Participant characteristics ($N = 212$)}
\label{tab:participant_characteristics}
\begin{tabular}{llrr}
\hline
\textbf{Characteristic} & \textbf{Category} & \textbf{$n$} & \textbf{\%} \\
\hline
Gender & Man & 120 & 56.6 \\
       & Woman & 48 & 22.6 \\
       & No response & 44 & 20.8 \\
\hline
Academic year & Second year & 7 & 3.3 \\
              & Third year & 66 & 31.1 \\
              & Fourth year & 124 & 58.5 \\
              & Unclassified entry & 15 & 7.1 \\
\hline
CGPA & Less than 2.50 & 9 & 4.2 \\
     & 2.50 to 3.00 & 54 & 25.5 \\
     & 3.00 to 3.25 & 37 & 17.5 \\
     & 3.25 to 3.50 & 50 & 23.6 \\
     & 3.50 to 3.75 & 38 & 17.9 \\
     & 3.75 to 4.00 & 24 & 11.3 \\
\hline
\end{tabular}
\vspace{0.5em}
\begin{minipage}{0.9\textwidth}
\small
\textit{Note.} Percentages may not sum to 100 because of rounding. No first-year respondents were present, as first-year students were not eligible. Academic year was recorded as free text and coded to year categories; unclassified entries are reported separately rather than distributed across years. CGPA categories are reproduced as presented to participants, with adjacent bands sharing boundary values.
\end{minipage}
\end{table}

\subsection{Procedure and Ethical Considerations}
\label{3.2}
Ethical approval for the study was granted by the Research Ethics Committee of the Faculty of Science and Information Technology, Daffodil International University (reference number IEC-FSIT/DIU/2025/2003; approval date 2 April 2025), under the protocol title “STEM Students’ Grumbling Attitude Towards Non-STEM Courses”.
The questionnaire was administered online through Google Forms between 10 April and 10 May 2025. The survey landing page presented an electronic participant information statement explaining the purpose of the research, eligibility requirements, an expected completion time of approximately 8 to 10 minutes, the voluntary nature of participation, foreseeable risks and benefits, data-handling arrangements, and the researchers’ contact details. Eligibility was restricted to students aged 18 years or older who were enrolled in an undergraduate Computer Science and Engineering programme and had completed at least one full academic year. Participants provided informed electronic consent by selecting the statement, “I have read and agree to statements 1–6 above and consent to take part,” before proceeding to the questionnaire. Those statements confirmed that the participant had read the information provided, met the eligibility criteria, understood that participation was voluntary and could be discontinued at any point before submission, understood that responses could not be withdrawn after submission because no identifying information was collected, and understood that the anonymous dataset would be published openly under a Creative Commons Attribution 4.0 licence. A separate optional item obtained permission for open-response comments to be published in the open dataset and quoted verbatim; participants could take part without agreeing to it. Respondents who did not provide consent were not permitted to continue. No incentives or payments of any kind were offered.
Participation was voluntary, and respondents could discontinue the survey before submission without penalty. The questionnaire did not collect names, student identification numbers, email addresses, IP addresses, or institutional affiliation, and it did not require participants to sign in. Participants were informed in advance that responses could not be withdrawn after submission because no direct identifiers were collected. Responses were assigned sequential, nonidentifying case numbers for analysis. Open-response comments were not analysed in this study and should remain publicly available only where optional publication consent can be documented, and disclosure control has been completed. A research copy will be retained for five years in accordance with the participant information statement, while the deidentified dataset published under a Creative Commons Attribution 4.0 licence remains publicly available.

\subsection{Instrument and Construct Operationalization}
\label{3.3}
\begin{landscape}
\renewcommand{\arraystretch}{1.15}
\setlength{\tabcolsep}{4pt}
\begin{longtable}{p{0.75cm} p{2.85cm} p{5.05cm} p{4.55cm} p{4.05cm} p{4.95cm}}
\caption{Provenance and adaptation of survey items relative to established measures}
\label{tab:item_provenance} \\
\hline
\textbf{No.} &
\textbf{Construct (Dimension)} &
\textbf{Study Item} &
\textbf{Source Instrument} &
\textbf{Original Item} &
\textbf{Nature of Adaptation} \\
\hline
\endfirsthead

\multicolumn{6}{l}{\textit{Table~\ref{tab:item_provenance} continued from previous page}} \\
\hline
\textbf{No.} &
\textbf{Construct (Dimension)} &
\textbf{Study Item} &
\textbf{Source Instrument} &
\textbf{Original Item} &
\textbf{Nature of Adaptation} \\
\hline
\endhead

\hline
\multicolumn{6}{r}{\textit{Continued on next page}} \\
\endfoot

\hline
\endlastfoot

1 &
Perceived academic burden (Cognitive Switching) &
PAB1 --- ``Adjusting my learning style for non-STEM courses is more challenging than for STEM courses.'' &
\cite{Dennis2010}. \textit{Cognitive Flexibility Inventory}. \textit{Cognitive Therapy and Research}, 34(3), 241--253. &
``I can find workable solutions to difficult situations.'' &
Domain specification: the source item assesses generalised adaptive flexibility across life situations; the present item restricts the referent to disciplinary learning-style adaptation between STEM and non-STEM coursework. \\

2 &
Perceived academic burden (Cognitive Switching) &
PAB2 --- ``Switching focus from STEM to non-STEM subjects is mentally exhausting.'' &
\cite{Hart}. NASA Task Load Index (NASA-TLX), Mental Demand subscale. In Hancock \& Meshkati (Eds.), \textit{Human Mental Workload} (pp. 139--183). North-Holland. &
``How mentally demanding was the task?'' &
Item-format conversion: the source item employs an interrogative rating-scale stem; the present item is restated as a declarative agreement statement and the referent narrowed from a generic task to interdisciplinary attentional switching. \\

3 &
Perceived academic burden (Workload Pressure) &
PAB3 --- ``Non-STEM courses significantly increase my academic workload.'' & \cite{Bedewy2015}. Perception of Academic Stress Scale (PAS), Workload and Examinations subscale. \textit{Health Psychology Open}, 2(2), 1--9. &
``The time allocated to classes and academic work is enough.'' &
Valence inversion and referent narrowing: the source item is a positively worded, reverse-scored sufficiency judgement applied to academic work in general; the present item is directly scored and attributes workload increase specifically to non-STEM coursework. \\

4 &
Perceived academic burden (Workload Pressure) &
PAB4 --- ``Managing both STEM and non-STEM courses is challenging.'' & \cite{Bedewy2015}. Perception of Academic Stress Scale (PAS). \textit{Health Psychology Open}, 2(2), 1--9. &
``The time allocated to classes and academic work is enough.'' &
Construct narrowing: the general workload-sufficiency judgement of the source instrument is recast as a bipartite management-difficulty judgement specific to concurrent STEM and non-STEM course loads. \\

5 &
Perceived academic burden (Work-Life Interference) &
PAB5 --- ``Balancing non-STEM and STEM coursework leaves me with less time for personal activities.'' & \cite{kopelman_model_1983}. Work--Family Conflict Scale, Time-Based Interference subscale. \textit{Organizational Behavior and Human Performance}, 32(2), 198--215. &
``My work takes up time that I'd like to spend with family/friends.'' &
Domain substitution: the occupational referent (``work'') is substituted with an academic referent (concurrent STEM/non-STEM coursework), and the personal-life referent is generalised from ``family/friends'' to ``personal activities.'' Structural parallelism with the source item is otherwise preserved. \\

6 &
Perceived benefits (Perceived Value) &
PB1 --- ``Non-STEM courses improve my soft skills, such as communication and critical thinking.'' & \cite{Byrne2020}. Attitudes Towards Learning Professional Skills Scale (ATLPS), Communication factor. \textit{Research in Science Education}. doi:10.1007/s11165-018-9738-3 &
``Developing good communication skills is just as important as mastering technical content in engineering.'' &
Construct-facet shift: an importance-appraisal item (relative value of communication versus technical content) is recast as an outcome-attribution item crediting non-STEM coursework with soft-skill development. \\

7 &
Perceived benefits (no label displayed) &
PB2 --- ``Studying non-STEM subjects broadens my overall perspective on learning.'' &
\cite{Byrne2020}. Attitudes Towards Learning Professional Skills Scale (ATLPS), Cultural Adaptability factor. \textit{Research in Science Education}. doi:10.1007/s11165-018-9738-3 &
``I think learning how to work with many nationalities and cultures is essential for my development as an engineer.'' &
Generalisation of referent: a specific cross-cultural competence item is broadened into a general claim of perspective-broadening attributable to non-STEM study. \\

8 &
Curricular devaluation (Role Ambiguity) &
RE1 --- ``I am unsure how non-STEM courses contribute to my academic goals.'' & \cite{Rizzo1970}. Role Ambiguity Scale. \textit{Administrative Science Quarterly}, 15(2), 150--163. &
``I know exactly what is expected of me.'' &
Domain transposition and valence inversion: the source item is a reverse-scored occupational role-clarity statement; the present item transposes the referent from organisational role to curricular purpose and is scored in the direction of ambiguity rather than clarity. \\

9 &
Curricular devaluation (Role Ambiguity) &
RE2 --- ``I struggle to see the practical application of non-STEM subjects compared to STEM subjects.'' &
\cite{Rizzo1970}. Role Ambiguity Scale. \textit{Administrative Science Quarterly}, 15(2), 150--163. &
``Clear, planned goals and objectives for my job.'' &
Domain transposition with added comparative framing: occupational goal clarity is reframed as perceived curricular utility, with an explicit STEM-versus-non-STEM comparison introduced that is absent from the source instrument. \\

10 &
Curricular devaluation (Negative Evaluative Orientation) &
RE3 --- ``I feel that non-STEM courses are a waste of my time compared to STEM courses.'' &
No close source item identified. The nearest conceptual context located was \cite{Byrne2020} discussion of engineering-student attitudes toward professional skills. &
No verbatim source item identified. &
Substantial new item: a direct comparative devaluation judgement. It should be treated as author-developed rather than as an adaptation of a published item or an empirical result. \\

11 &
Curricular devaluation (no label displayed) &
RE4 --- ``Courses in the humanities (e.g., literature, philosophy) are unnecessary for STEM students.'' & \cite{Byrne2020}. Item reported during ATLPS scale development but not retained in the final scale. \textit{Research in Science Education}. &
Closest source wording: ``communication skills \ldots{} not required of engineering students.'' &
Domain generalisation with reverse-scoring removed: the source item was not retained in the published scale and only partially anticipates the present item. ``Communication skills'' is generalised to humanities courses; the item should therefore be treated as a substantial adaptation. \\

12 &
Curricular devaluation (no label displayed) &
RE5 --- ``Social science courses (e.g., sociology, psychology) seem less relevant to my field than STEM subjects.'' &
No close source item identified. The nearest conceptual context located was \cite{Byrne2020} discussion of engineering-student attitudes toward professional skills. &
No verbatim source item identified. &
Substantial new item: a comparative relevance judgement narrowed to social science coursework. It should be treated as author-developed rather than as an adaptation of a published item or an empirical result. \\

13 &
Integrative intention (Knowledge Integration and Application) &
PO1 --- ``I believe non-STEM courses are valuable despite their challenges.'' & \cite{eccles2020expectancy}. Expectancy--Value Theory, cost component of subjective task value. &
Theoretical construct (no single verbatim item): perceived task value net of perceived cost (time, effort, difficulty). &
Operationalisation of a theoretical construct: the item translates the expectancy-value ``value-despite-cost'' proposition into a direct self-report statement; no corresponding published item was located. \\

14 &
Integrative intention (Knowledge Integration and Application) &
PO2 --- ``I am open to integrating lessons from non-STEM courses into my academic work.'' & \cite{Schijf2023}. Interdisciplinary Understanding Questionnaire (IUQ), Knowledge of Interdisciplinarity subscale. \url{https://doi.org/10.1080/21568235.2022.2058045} &
``While solving an academic problem, I am good at figuring out which information from outside my own field of study I can use.'' &
Construct-facet shift: a self-efficacy item (perceived competence at cross-domain integration) is recast as a dispositional-openness item (willingness to integrate), retaining the underlying integration theme. \\

15 &
Integrative intention (Knowledge Integration and Application) &
PO3 --- ``I will likely apply the knowledge I gain from non-STEM courses in my personal or professional life.'' &
(Pintrich et al., 1991). Motivated Strategies for Learning Questionnaire (MSLQ), Task Value subscale. &
``I think I will be able to use what I learn in this course in other courses.'' &
Scope broadening and modality shift: the transfer referent is broadened from ``other courses'' to ``personal or professional life,'' and the epistemic framing shifts from perceived ability (``I think I will be able to'') to behavioural likelihood (``I will likely apply''). \\

\end{longtable}

\vspace{0.5em}
\noindent\begin{minipage}{22.2cm}
\small
\textit{Note.} Item wording is reproduced exactly as displayed to participants. Parenthetical terms are the subconstruct labels displayed beside 12 items. Three items carried no label. Codes are the analysis codes used in the manuscript. Rows are ordered by construct rather than fielding order. The measures are study-specific adaptations and are not presented as fully validated scales.
\end{minipage}
\end{landscape}

The construct labels were determined by the substantive content of the items rather than adopted uncritically from the terminology used in the original instrument. The five items coded RE1 to RE5 do not measure protest, avoidance, oppositional behaviour, or affective resistance. Instead, they assess uncertainty about the academic contribution and practical application of non-STEM coursework, together with judgements that such coursework is unnecessary, less relevant than technical subjects, or a poor use of time. These indicators were therefore interpreted as measuring curricular devaluation.
Similarly, the three items coded PO1 to PO3 do not measure observed acceptance, engagement, knowledge transfer, or subsequent professional behaviour. They assess perceived value despite difficulty, openness to incorporating broader learning into academic work, and anticipated future application. These indicators were therefore operationalised as integrative intention. This terminology aligns the construct claims more closely with what the questionnaire actually measured.
One feature of the fielded instrument requires disclosure. Twelve of the 15 items displayed the originally intended subconstruct label in parentheses after the statement, including ‘Cognitive Switching’, ‘Perceived Value’, ‘Workload Pressure’, ‘Role Ambiguity’, ‘Work-Life Interference’, ‘Negative Evaluative Orientation’, and ‘Knowledge Integration and Application’. These labels reflect the narrower taxonomy around which the instrument was initially written, whereas the present analysis specifies four constructs on the basis of item content. Because visible labels can cue internally consistent responding within a labelled group, they may contribute to the observed within-construct correlations and the performance of the four-factor measurement model. The labels are reproduced in Table~\ref{tab:item_provenance}, and this issue is treated as a limitation in Section~\ref{sec:limitations}.
Perceived academic burden, curricular devaluation, and integrative intention were specified as latent variables. Perceived benefits were measured by only two items. Although a two-indicator latent benefits factor was examined in the measurement and sensitivity analyses, its nonparametric resampling performance was unstable. The primary structural model therefore represented perceived benefits using the mean of PB1 and PB2 as an observed index. This modelling decision is explained further in Section~\ref{3.5}. Table~\ref{tab:construct_operationalisation} summarises the content, indicator mapping, and analytical role of each construct included in the proposed model.

\begin{table}[htbp]
\centering
\caption{Operationalisation of the study constructs}
\label{tab:construct_operationalisation}
\renewcommand{\arraystretch}{1.2}
\setlength{\tabcolsep}{5pt}
\begin{tabular}{p{0.18\textwidth} p{0.12\textwidth} p{0.36\textwidth} p{0.25\textwidth}}
\hline
\textbf{Construct} & \textbf{Codes} & \textbf{Content represented} & \textbf{Primary model role} \\
\hline
Perceived academic burden &
PAB1 to PAB5 &
Cognitive switching, workload pressure, difficulty coordinating technical and non-STEM coursework, and interference with personal time. &
Latent exogenous construct \\
\hline
Perceived benefits &
PB1 to PB2 &
Development of communication and critical thinking skills and a broader perspective on learning. &
Observed mean index in the primary structural model; latent factor examined in the measurement and sensitivity analyses \\
\hline
Curricular devaluation &
RE1 to RE5 &
Uncertainty about academic contribution and practical application, perceived waste of time, lack of necessity, and lower relevance relative to technical subjects. &
Latent endogenous construct positioned as the evaluative link between academic appraisals and integrative intention \\
\hline
Integrative intention &
PO1 to PO3 &
Perceived value despite difficulty, openness to integration, and anticipated application in later academic, personal, or professional contexts. &
Latent endogenous outcome construct \\
\hline
\end{tabular}
\end{table}

\noindent Table~\ref{tab:item_provenance} documents the source instrument, source wording, and nature of adaptation for every item. No formal expert content review, pilot study, or cognitive interviewing was conducted before fielding. The measures are therefore reported as study-specific operationalisations rather than fully validated scales, and their psychometric properties should be treated as provisional \cite{Boateng2018}.

The questionnaire was administered entirely in English; no translated version or back-translation procedure was used. Of 636 possible entries across the three open-response fields, 630 were nonblank. The nonnumeric entries used Latin script, no Bengali script was present, and two entries consisted solely of numerals. English-only administration was a design choice rather than evidence of linguistic equivalence: the instrument was not cognitively pretested across different levels of English proficiency.

\subsection{Data Preparation and Response Quality}
\label{3.4}
The dataset contained 212 cases with complete responses across all 15 focal variables, so no imputation was required. All five response categories were observed for every item. Screening identified 204 unique response vectors and eight exact duplicate vectors beyond the first occurrence. Ten respondents gave invariant responses across all 15 items. Using the stricter longstring definition of an uninterrupted run of identical responses in fielding order, 12 respondents reached a run of 12 or more items. In a separate dominant category check, 21 respondents used one response category for at least 12 of the 15 items. Exact duplication, invariant responding, long strings, and dominant category use can indicate insufficient effort, but none is conclusive when response time, instructed response items, attention checks, and process data are unavailable \cite{meade_identifying_2012}.
The full sample was retained for the primary analysis. Two sensitivity samples were used to determine whether highly patterned responding materially influenced the conclusions. The first excluded the ten invariant response vectors, producing n = 202. The second excluded the 21 respondents whose dominant response category accounted for at least 12 items, producing n = 191. Section~\ref{sec:sensitivity} reports these analyses. Scale scores used for descriptive reporting were arithmetic means of the assigned items, and the mean of PB1 and PB2 represented perceived benefits in the primary structural model. No observation was removed because of the direction or extremity of its substantive responses, and no post hoc residual covariance was added to improve model fit.

\subsection{Analytical Strategy}
\label{3.5}
Analyses were implemented in R using the lavaan package \cite{Rosseel2012}. The final analysis used random seed 20260807 to support reproducible bootstrap estimation. All latent variables were identified by fixing their variances to one, and the observed benefits index was treated as a random exogenous variable.
The analysis followed the two-step latent variable modelling strategy proposed by \cite{Anderson1988}. In the first step, confirmatory factor analysis was used to evaluate the hypothesised four-factor measurement structure. The proposed model was compared with a one-factor model and three theoretically relevant alternatives in which conceptually adjacent constructs were combined. All measurement models were estimated using robust maximum likelihood, MLR. Measurement quality was evaluated using standardised factor loadings, Cronbach’s alpha, composite reliability, average variance extracted, latent construct correlations, and the heterotrait-monotrait ratio, HTMT. HTMT was computed using the arithmetic mean formulation proposed by \cite{Henseler2015}, and values below .85 were treated as evidence consistent with discriminant validity. Reliability and convergent validity were evaluated collectively rather than determined through a single mechanical threshold \cite{Henseler2015, Kline}.
In the second step, the structural component specified two linked regression equations. For respondent (i), the structural relations were expressed as:

\begin{equation}
CD_i = \alpha_1 + \beta_1 PAB_i + \beta_2 PB_i + \varepsilon_{1i}
\tag{1}
\end{equation}

\begin{equation}
II_i = \alpha_2 + \beta_3 CD_i + \beta_4 PAB_i + \beta_5 PB_i + \varepsilon_{2i}
\tag{2}
\end{equation}

where CD denotes curricular devaluation, PAB denotes perceived academic burden, PB denotes the observed perceived benefits index, and II denotes integrative intention for respondent i. The terms α₁ and α₂ are equation intercepts, β₁ to β₅ are structural coefficients, and ε₁ᵢ and ε₂ᵢ are residual terms.
The indirect association between academic burden and integrative intention through curricular devaluation was calculated as β₁β₃, corresponding to H6a. The corresponding indirect association for perceived benefits was calculated as β₂β₃, corresponding to H6b. The total association for academic burden was β₄ + β₁β₃, and the total association for perceived benefits was β₅ + β₂β₃. Perceived academic burden, curricular devaluation, and integrative intention were represented as latent variables, while perceived benefits were represented by the arithmetic mean of PB1 and PB2. Mardia’s multivariate skewness and kurtosis tests indicated substantial departure from multivariate normality \cite{MARDIA1970}. The primary model was therefore estimated using MLR with robust standard errors and a scaled test statistic.
Model fit was evaluated using the scaled chi-square statistic, robust comparative fit index, robust Tucker–Lewis index, robust root mean square error of approximation with its 90\% confidence interval, and standardised root mean square residual. These indices were interpreted collectively and in relation to model complexity, measurement quality, and theoretically plausible alternatives rather than through a rigid pass or fail rule. The original structural specification represented perceived benefits as a latent variable measured by PB1 and PB2. This two-indicator specification produced unstable nonparametric resampling. Of 5,000 bootstrap draws, 24 failed or did not converge and 1,034 produced nonadmissible solutions, so approximately one draw in five could not be used. The instability is expressed in this rate of inadmissibility rather than in the dispersion of the usable estimates, which was unremarkable among the draws that did converge to an admissible solution. The primary structural model therefore used the mean of PB1 and PB2 as an observed benefits index. Estimated under the same 5,000 resamples, this specification produced no failed or non-converged draws, fewer than one nonadmissible draw in a thousand, and no estimation warnings. The latent benefits specification was retained as a sensitivity analysis. Although using an observed index does not correct explicitly for measurement error, it avoids basing the indirect effect analysis on an empirically unstable two-indicator latent factor.
Indirect and total associations were evaluated using 5,000 respondent-level nonparametric bootstrap resamples. Because respondent-level nonparametric bootstrap estimation was not conducted directly under MLR, the bootstrap draws were generated from an otherwise identical model estimated using maximum likelihood. The primary direct-path estimates, robust standard errors, confidence intervals, and model fit statistics remained based on MLR. Percentile bootstrap confidence intervals were presented for the indirect and total associations because products of coefficients commonly have asymmetric sampling distributions. All 5,000 resamples converged and produced admissible estimates. The necessity of the two direct appraisals to intention pathways was evaluated by comparing the primary partial model with an indirect-only model in which both direct pathways were constrained to zero. The nested models were compared using a Satorra–Bentler scaled chi-square difference test. Because a non-significant coefficient does not establish that a pathway is absent, each direct pathway was additionally evaluated on its own by comparing the primary model with a model constraining only the burden to intention pathway to zero and with a model constraining only the benefits to intention pathway to zero.
Three sensitivity analyses were conducted. First, the structural model was re-estimated using the original two-indicator latent benefits factor. Second, the primary model was re-estimated after excluding the ten invariant response vectors. Third, it was re-estimated after excluding respondents for whom a single response category accounted for at least 12 of the 15 items. Coefficient direction, magnitude, confidence intervals, and hypothesis decisions were compared across these specifications.
As an estimator sensitivity analysis, the model was re-estimated using the weighted least squares mean- and variance-adjusted estimator \cite{Li2016}. The 13 indicators measuring academic burden, curricular devaluation, and integrative intention were declared ordered, while the mean of PB1 and PB2 was retained as an observed exogenous benefits index so that the ordinal specification remained comparable with the primary MLR model. The ordinal model used theta parameterization, and four thresholds were estimated for each ordered indicator. Because respondent-level nonparametric resampling is not well defined for the polychoric input used by this estimator, confidence intervals for the indirect and total associations were obtained by the delta method rather than by bootstrap resampling and are therefore not directly comparable with the percentile bootstrap intervals described for the primary model. Because the mean- and variance-adjusted test statistic is both scaled and shifted, the robust fit indices defined for maximum likelihood estimation are undefined for this specification, so scaled comparative fit index, scaled Tucker–Lewis index, and scaled root mean square error of approximation values are stated instead.
The structural component was saturated conditional on the measurement model. Consequently, global fit primarily reflects the adequacy of the latent measurement structure and cannot establish the direction, temporal sequence, or causal status of the structural pathways. Because the data were cross-sectional, the direct, indirect, and total estimates are interpreted as associations consistent with the theoretically specified model rather than as evidence of causal mediation.

\section{Result}

The results are presented in four stages. Section~\ref{4.1} presents descriptive statistics and distributional diagnostics. Section~\ref{4.2} evaluates the measurement structure and construct quality. Section~\ref{4.3} presents the direct structural associations, and Section~\ref{sec:sensitivity} examines the indirect, total, and sensitivity estimates.

\subsection{Descriptive Statistics and Distributional Diagnostics}
\label{4.1}
Item means ranged from 3.12 to 3.79, and every response category was observed for every indicator. Mardia's tests indicated significant multivariate skewness, 40.35, $\chi^2(680) = 1425.57$, $p < .001$, and multivariate kurtosis, 308.69, $z = 17.31$, $p < .001$ \cite{MARDIA1970}. These departures from multivariate normality supported the use of robust estimation and inference.

Table~\ref{tab:descriptive_reliability_correlations} summarises the construct means, standard deviations, internal consistency estimates, and correlations among the scale scores.

\begin{table}[htbp]
\centering
\caption{Descriptive statistics, reliability, and scale score correlations}
\label{tab:descriptive_reliability_correlations}
\renewcommand{\arraystretch}{1.15}
\setlength{\tabcolsep}{6pt}
\begin{tabular}{lrrrrrrr}
\hline
\textbf{Construct} & \textbf{$M$} & \textbf{$SD$} & \textbf{$\alpha$} &
\textbf{1} & \textbf{2} & \textbf{3} & \textbf{4} \\
\hline
1. Perceived academic burden & 3.60 & .66 & .770 & 1.000 & & & \\
2. Perceived benefits & 3.47 & .81 & .672 & $-.078$ & 1.000 & & \\
3. Curricular devaluation & 3.36 & .77 & .781 & .519 & $-.266$ & 1.000 & \\
4. Integrative intention & 3.35 & .65 & .681 & $-.147$ & .386 & $-.398$ & 1.000 \\
\hline
\end{tabular}
\vspace{0.5em}
\begin{minipage}{0.92\textwidth}
\small
\textit{Note.} Means ($M$) and standard deviations ($SD$) were calculated from unweighted item means coded from 1 to 5. Correlations are Pearson correlations among the scale scores. Because perceived benefits was measured using only two items, its alpha coefficient should be interpreted cautiously.
\end{minipage}
\end{table}

The correlation pattern was consistent with the proposed model. Academic burden was positively related to curricular devaluation, while benefits were negatively related to devaluation and positively related to integrative intention. The modest negative bivariate association between burden and intention also indicates that burden and intention were not simply opposite ends of a single evaluative continuum.

\subsection{Measurement Model}
\label{4.2}

The hypothesised four-factor measurement structure was compared with a one-factor model and three theoretically plausible collapsed-factor models. Table~\ref{tab:measurement_model_comparisons} presents the comparative fit statistics for these specifications.

\begin{table}[htbp]
\centering
\caption{Measurement model comparisons using robust maximum likelihood}
\label{tab:measurement_model_comparisons}
\renewcommand{\arraystretch}{1.15}
\setlength{\tabcolsep}{3pt}
\footnotesize
\begin{tabular}{>{\raggedright\arraybackslash}p{0.24\textwidth} >{\centering\arraybackslash}p{0.115\textwidth} >{\centering\arraybackslash}p{0.05\textwidth} >{\centering\arraybackslash}p{0.10\textwidth} >{\centering\arraybackslash}p{0.10\textwidth} >{\centering\arraybackslash}p{0.19\textwidth} >{\centering\arraybackslash}p{0.075\textwidth}}
\hline
\textbf{Model} &
\textbf{Scaled $\chi^2$} &
\textbf{df} &
\textbf{Robust CFI} &
\textbf{Robust TLI} &
\textbf{Robust RMSEA [90\% CI]} &
\textbf{SRMR} \\
\hline
Four factors &
149.03 & 84 & .911 & .888 & .066 [.048, .082] & .065 \\
One factor &
339.30 & 90 & .646 & .587 & .126 [.112, .140] & .111 \\
Burden and devaluation merged &
226.71 & 87 & .808 & .769 & .094 [.079, .109] & .086 \\
Benefits and intention merged &
176.61 & 87 & .875 & .849 & .076 [.060, .092] & .071 \\
Devaluation and intention merged &
206.35 & 87 & .833 & .798 & .088 [.073, .104] & .085 \\
\hline
\end{tabular}
\vspace{0.5em}
\begin{minipage}{0.94\textwidth}
\small
\textit{Note.} RMSEA values are robust. SRMR is unscaled. No post hoc residual covariances were added.
\end{minipage}
\end{table}

The hypothesised four-factor model fitted better than every alternative specification examined, but its absolute fit was mixed rather than uniformly good. Robust CFI (.911) and SRMR (.065) fell within conventionally acceptable ranges, whereas robust TLI (.888) fell below .90 and the upper bound of the robust RMSEA confidence interval (.082) exceeded the .08 value commonly treated as the limit of acceptable approximate fit. Each collapsed model fit materially worse, and the one-factor model showed substantial misfit. The measurement evidence therefore supports the four-factor structure comparatively, as the best available representation among the theoretically plausible alternatives, rather than absolutely. It should not be described as demonstrating good or adequate fit.

Table~\ref{tab:measurement_quality} describes the standardised loading ranges and the principal reliability and convergent validity indicators for each construct.

\begin{table}[htbp]
\centering
\caption{Measurement quality of the study constructs}
\label{tab:measurement_quality}
\renewcommand{\arraystretch}{1.15}
\setlength{\tabcolsep}{3pt}
\footnotesize
\begin{tabular}{>{\raggedright\arraybackslash}p{0.17\textwidth} >{\centering\arraybackslash}p{0.055\textwidth} >{\centering\arraybackslash}p{0.15\textwidth} >{\centering\arraybackslash}p{0.05\textwidth} >{\centering\arraybackslash}p{0.13\textwidth} >{\centering\arraybackslash}p{0.055\textwidth} >{\raggedright\arraybackslash}p{0.22\textwidth}}
\hline
\textbf{Construct} &
\textbf{Items} &
\textbf{Standardised loading range} &
\textbf{$\alpha$} &
\textbf{Composite reliability} &
\textbf{AVE} &
\textbf{Assessment} \\
\hline
Perceived academic burden &
5 &
.561 to .764 &
.770 &
.777 &
.413 &
Reliability acceptable; convergent validity modest \\
\hline
Perceived benefits &
2 &
.601 to .846 &
.672 &
.695 &
.539 &
AVE adequate; two-item reliability requires caution \\
\hline
Curricular devaluation &
5 &
.563 to .738 &
.781 &
.782 &
.420 &
Reliability acceptable; convergent validity modest \\
\hline
Integrative intention &
3 &
.597 to .722 &
.681 &
.687 &
.424 &
Reliability and convergent validity modest \\
\hline
\end{tabular}
\vspace{0.5em}
\begin{minipage}{0.94\textwidth}
\small
\textit{Note.} The benefits loading range, composite reliability, and AVE were obtained from the latent benefits measurement specification. Perceived benefits was represented by the mean of PB1 and PB2 in the primary structural model. AVE denotes average variance extracted.
\end{minipage}
\end{table}

Standardised loadings were positive and substantively aligned with their assigned constructs. HTMT values ranged from .199 to .673, remaining below the .85 criterion and supporting discriminant validity. Nevertheless, AVE was below .50 for burden, curricular devaluation, and integrative intention, while reliability for the benefits and intention measures was modest. The evidence therefore supports empirical differentiation among the constructs but does not establish definitive scale validation.

\subsection{Structural Associations}
\label{4.3}
The primary structural model showed mixed fit: scaled $\chi^2(72) = 128.01$, $p < .001$, robust CFI = .914, robust TLI = .891, robust RMSEA = .067 with a 90\% confidence interval from .047 to .085, and SRMR = .061. Robust CFI and SRMR fell within conventionally acceptable ranges, whereas robust TLI fell below .90 and the upper bound of the robust RMSEA confidence interval exceeded .08, so the structural model fit is likewise described as mixed. It explained 47.6\% of the variance in curricular devaluation and 40.6\% of the variance in integrative intention.

Table~\ref{tab:structural_estimates} presents the robust standardised estimates for the five hypothesised direct associations.

\begin{table}[htbp]
\centering
\caption{Robust standardised structural estimates}
\label{tab:structural_estimates}
\renewcommand{\arraystretch}{1.15}
\setlength{\tabcolsep}{4pt}
\begin{tabularx}{\textwidth}{l X r r c c l}
\hline
\textbf{Hypothesis} &
\textbf{Structural path} &
\textbf{$\beta$} &
\textbf{Robust SE} &
\textbf{95\% robust CI} &
\textbf{$p$} &
\textbf{Decision} \\
\hline
H1 & Burden $\rightarrow$ devaluation & .620 & .070 & [.482, .758] & $< .001$ & Supported \\
H2 & Benefits $\rightarrow$ devaluation & $-.250$ & .063 & [$-.373$, $-.127$] & $< .001$ & Supported \\
H3 & Devaluation $\rightarrow$ intention & $-.565$ & .131 & [$-.821$, $-.309$] & $< .001$ & Supported \\
H4 & Burden $\rightarrow$ intention & .177 & .138 & [$-.093$, .447] & .200 & Not supported \\
H5 & Benefits $\rightarrow$ intention & .299 & .085 & [.132, .466] & $< .001$ & Supported \\
\hline
\end{tabularx}
\vspace{0.5em}
\begin{minipage}{0.94\textwidth}
\small
\textit{Note.} $\beta$ denotes the fully standardised estimate. Confidence intervals were calculated using robust sandwich standard errors under MLR. H4 predicted a negative residual association, but the estimated coefficient was positive and its confidence interval included zero.
\end{minipage}
\end{table}

Four of the five direct hypotheses were supported. Academic burden showed a strong positive association with curricular devaluation, while perceived benefits showed a negative association with devaluation. Curricular devaluation, in turn, was negatively associated with integrative intention. Benefits also retained a positive association with intention after burden and devaluation were included in the model. The direct burden to intention pathway was not supported. Its positive point estimate was opposite to the hypothesised direction, but the confidence interval included zero. This result should be described as an unsupported direct association rather than evidence that burden increases intention. Under the ordinal estimator presented in Section~\ref{sec:sensitivity}, however, this pathway did reach conventional significance in the same positive direction, so its status is estimator dependent. A model constraining direct appraisal to intention pathways to zero fit worse than the partial model, scaled difference $\chi^2(2) = 14.40$, $p < .001$. The joint test indicates that at least one direct appraisal pathway was required. Considered together with the path-specific estimates, the difference was attributable principally to the positive benefits to intention association rather than the unsupported burden pathway.

Because a non-significant coefficient does not establish that a pathway is absent, each direct appraisal-to-intention pathway was additionally evaluated through a separate nested comparison presented in Table~\ref{tab:nested_direct_pathways}.

\begin{sidewaystable}[htbp]
\centering
\caption{Nested comparison of the direct appraisal-to-intention pathways}
\label{tab:nested_direct_pathways}
\renewcommand{\arraystretch}{1.15}
\setlength{\tabcolsep}{2.5pt}
\small
\begin{tabular}{>{\raggedright\arraybackslash}p{0.135\textheight} >{\centering\arraybackslash}p{0.068\textheight} >{\centering\arraybackslash}p{0.036\textheight} >{\centering\arraybackslash}p{0.072\textheight} >{\centering\arraybackslash}p{0.135\textheight} >{\centering\arraybackslash}p{0.054\textheight} >{\centering\arraybackslash}p{0.072\textheight} >{\centering\arraybackslash}p{0.072\textheight} >{\centering\arraybackslash}p{0.126\textheight} >{\centering\arraybackslash}p{0.063\textheight}}
\hline
\textbf{Model} &
\textbf{Scaled $\chi^2$} &
\textbf{df} &
\textbf{Robust CFI} &
\textbf{Robust RMSEA [90\% CI]} &
\textbf{SRMR} &
\textbf{AIC} &
\textbf{BIC} &
\textbf{Scaled $\Delta\chi^2$ ($\Delta$df)} &
\textbf{$p$} \\
\hline
Primary (both direct pathways free) &
128.01 & 72 & .914 & .067 [.047, .085] & .061 &
7324.80 & 7435.56 & --- & --- \\
Burden to intention fixed to zero &
129.57 & 73 & .913 & .066 [.047, .085] & .063 &
7324.73 & 7432.14 & 1.56 (1) & .211 \\
Benefits to intention fixed to zero &
139.43 & 73 & .899 & .072 [.053, .090] & .065 &
7335.60 & 7443.01 & 18.30 (1) & $< .001$ \\
Both direct pathways fixed to zero &
143.10 & 74 & .894 & .073 [.055, .091] & .071 &
7339.32 & 7443.37 & 14.40 (2) & $< .001$ \\
\hline
\end{tabular}
\vspace{0.5em}
\begin{minipage}{0.9\textheight}
\small
\textit{Note.} Each constrained model is compared with the primary model using the Satorra--Bentler scaled difference test. The scaled difference statistic is not additive across constraints, so the joint two-pathway test does not equal the sum of the two single-pathway tests. AIC and BIC are computed from the maximum likelihood fit function. $R^2$ is not used as a comparison criterion because constraining a direct pathway inflates the devaluation-to-intention coefficient, which holds explained variance artificially high.
\end{minipage}
\end{sidewaystable}

Constraining the burden to intention pathway to zero did not produce a significant deterioration in fit, scaled $\Delta\chi^2(1) = 1.56$, $p = .211$, and left the robust comparative fit index essentially unchanged, from .914 to .913, while both the Akaike and Bayesian information criteria marginally favoured the constrained model. Constraining the benefits to intention pathway to zero produced a significant deterioration, scaled $\Delta\chi^2(1) = 18.30$, $p < .001$, together with a decline in robust CFI to .899 and an increase in robust RMSEA to .072, and both information criteria favoured retaining the pathway. The two direct pathways therefore contribute differently to model fit. This pattern indicates different pathway configurations for burden and benefits rather than demonstrating that the burden to intention association is zero. The standardised estimate for that pathway was .177 with a 95\% confidence interval from $-.093$ to .447, an interval too wide to support a claim of equivalence to zero, and the present design cannot distinguish a genuinely absent association from one too small to be detected at this sample size.

\begin{figure}[htbp]
\centering
\includegraphics[width=0.9\textwidth]{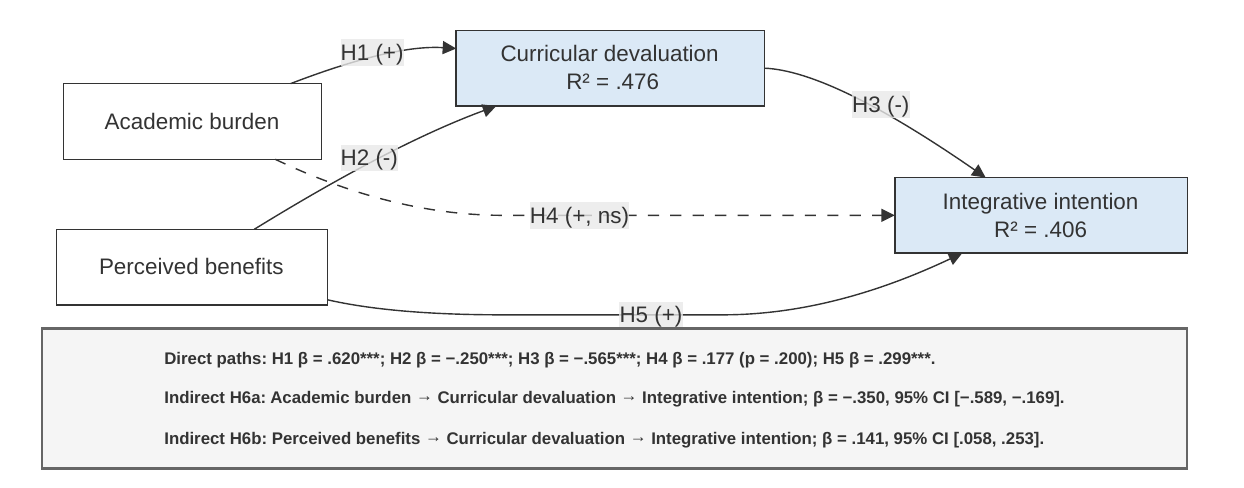}
\caption{Primary structural model with robust standardised estimates}
\label{fig:primary_structural_model}
\vspace{0.5em}
\end{figure}

Figure~\ref{fig:primary_structural_model} displays the estimated structural model, standardised coefficients, and explained variance in the two endogenous constructs.

\subsection{Indirect, Total, and Sensitivity Estimates}
\label{sec:sensitivity}
The theoretically specified indirect and total associations were evaluated using 5,000 respondent-level bootstrap resamples. Table~\ref{tab:bootstrap_associations} reports the standardised estimates, bootstrap uncertainty, and percentile confidence intervals.

\begin{table}[htbp]
\centering
\caption{Bootstrap indirect and total associations}
\label{tab:bootstrap_associations}
\renewcommand{\arraystretch}{1.15}
\setlength{\tabcolsep}{5pt}
\begin{tabularx}{\textwidth}{X r r c c}
\hline
\textbf{Association} &
\textbf{$\beta$} &
\textbf{Bootstrap SE} &
\textbf{95\% percentile CI} &
\textbf{Empirical $p$} \\
\hline
Burden $\rightarrow$ devaluation $\rightarrow$ intention &
$-.350$ & .108 & [$-.589$, $-.169$] & $< .001$ \\
Burden $\rightarrow$ intention, total &
$-.174$ & .113 & [$-.378$, .065] & .148 \\
Benefits $\rightarrow$ devaluation $\rightarrow$ intention &
.141 & .050 & [.058, .253] & $< .001$ \\
Benefits $\rightarrow$ intention, total &
.440 & .082 & [.276, .600] & $< .001$ \\
\hline
\end{tabularx}
\vspace{0.5em}
\begin{minipage}{0.94\textwidth}
\small
\textit{Note.} Estimates are standardised products from 5,000 respondent-level bootstrap resamples. All resamples converged in the observed benefits index specification.
\end{minipage}
\end{table}

The indirect association between burden and intention through curricular devaluation was negative, supporting H6a. However, the total burden to intention association was not distinguishable from zero. This combination indicates that the data support the theoretically specified indirect association but do not support a general claim that burden was associated with lower intention across all represented pathways. Benefits showed a positive indirect association with intention through lower curricular devaluation, supporting H6b. Benefits also retained a positive direct association and consequently had a positive total association with intention. The different direct and indirect configurations of burden and benefits provide the empirical basis for describing the appraisal structure as asymmetric. Sensitivity analyses examined whether the structural pattern depended on the observed benefits index or on respondents showing highly patterned answers.

Table~\ref{tab:sensitivity_structural_estimates} compares the standardised coefficients across the primary and alternative specifications.

\begin{sidewaystable}[htbp]
\centering
\caption{Sensitivity of standardised structural estimates}
\label{tab:sensitivity_structural_estimates}
\renewcommand{\arraystretch}{1.15}
\setlength{\tabcolsep}{5pt}
\small
\begin{tabular}{>{\raggedright\arraybackslash}p{0.16\textheight} >{\centering\arraybackslash}p{0.035\textheight} >{\centering\arraybackslash}p{0.13\textheight} >{\centering\arraybackslash}p{0.13\textheight} >{\centering\arraybackslash}p{0.13\textheight} >{\centering\arraybackslash}p{0.13\textheight} >{\centering\arraybackslash}p{0.13\textheight}}
\hline
\textbf{Specification} &
\textbf{$n$} &
\textbf{Burden $\rightarrow$ devaluation} &
\textbf{Benefits $\rightarrow$ devaluation} &
\textbf{Devaluation $\rightarrow$ intention} &
\textbf{Burden $\rightarrow$ intention} &
\textbf{Benefits $\rightarrow$ intention} \\
\hline
Primary observed benefits index &
212 & .620 & $-.250$ & $-.565$ & .177 ns & .299 \\
Latent two-item benefits factor &
212 & .592 & $-.304$ & $-.518$ & .175 ns & .342 \\
Invariant vectors excluded &
202 & .604 & $-.258$ & $-.596$ & .155 ns & .265 \\
Dominant response category used for at least 12 items excluded &
191 & .595 & $-.255$ & $-.640$ & .166 ns & .255 \\
\hline
\end{tabular}
\vspace{0.5em}
\begin{minipage}{0.9\textheight}
\small
\textit{Note.} All coefficients are standardised. The latent benefits model is retained only as a specification sensitivity because its bootstrap analysis generated nonadmissible solutions. Across specifications, ns indicates that the confidence interval for the burden-to-intention pathway included zero.
\end{minipage}
\end{sidewaystable}

The coefficient signs and substantive hypothesis decisions remained stable across the alternative specifications. The devaluation to intention association became somewhat stronger after patterned response vectors were excluded, while the direct burden pathway remained unsupported. These analyses do not eliminate the study's measurement and sampling limitations, but they indicate that the central asymmetric pattern was not attributable solely to the observed benefits index or to a small group of highly patterned respondents.

The ordinal specification converged to an admissible solution with no estimation warnings. Its fit was mixed and weaker in approximation terms than the primary model: scaled $\chi^2(72) = 188.58$, $p < .001$, scaled CFI = .920, scaled TLI = .898, scaled RMSEA = .088 with a 90\% confidence interval from .072 to .103, SRMR = .065, and WRMR = .969. Standardised coefficients were consistent in direction with the primary model and somewhat larger in magnitude, and explained variance rose to 53.9\% for curricular devaluation and 42.4\% for integrative intention. The indirect pathways were unchanged in substance, with burden retaining a negative indirect association through devaluation and benefits a positive one, both with intervals excluding zero. Two estimates diverged. The direct burden to intention pathway, unsupported under MLR, reached conventional significance under the ordinal estimator, $\beta = .232$, 95\% CI [.016, .448], $p = .035$, in a positive direction still opposite to that predicted by H4. The total burden to intention association, not distinguishable from zero under MLR, was negative and significant under the ordinal estimator, $\beta = -.180$, 95\% CI [$-.323$, $-.036$], $p = .014$. The asymmetric pathway configuration therefore held under ordinal estimation, but the conclusion that burden carries no additional direct association with intention is estimator dependent. The combination of a positive direct and a negative indirect association is an inconsistent direct-indirect pattern in which the two routes partially offset one another. It is not evidence of causal mediation and is reported here as a boundary on the strength of the asymmetry claim rather than as a competing result.

The complete MLR--WLSMV comparison is presented in Table~\ref{tab:mlr_wlsmv_comparison}.

\begin{table}[htbp]
\centering
\caption{Ordinal WLSMV estimator sensitivity analysis}
\label{tab:mlr_wlsmv_comparison}
\renewcommand{\arraystretch}{1.15}
\setlength{\tabcolsep}{5pt}
\begin{tabular}{p{0.28\textwidth} c c c c}
\hline
\textbf{Parameter} &
\textbf{MLR $\beta$} &
\textbf{MLR 95\% CI} &
\textbf{WLSMV $\beta$} &
\textbf{WLSMV 95\% CI} \\
\hline
Burden $\rightarrow$ devaluation &
.620 & [.482, .758] & .673 & [.588, .759] \\
Benefits $\rightarrow$ devaluation &
$-.250$ & [$-.373$, $-.127$] & $-.237$ & [$-.328$, $-.147$] \\
Devaluation $\rightarrow$ intention &
$-.565$ & [$-.821$, $-.309$] & $-.612$ & [$-.808$, $-.416$] \\
Burden $\rightarrow$ intention &
.177 ns & [$-.093$, .447] & .232 & [.016, .448] \\
Benefits $\rightarrow$ intention &
.299 & [.132, .466] & .306 & [.197, .416] \\
Burden $\rightarrow$ devaluation $\rightarrow$ intention &
$-.350$ & [$-.589$, $-.169$] & $-.412$ & [$-.569$, $-.254$] \\
Benefits $\rightarrow$ devaluation $\rightarrow$ intention &
.141 & [.058, .253] & .145 & [.073, .217] \\
Burden $\rightarrow$ intention, total &
$-.174$ ns & [$-.378$, .065] & $-.180$ & [$-.323$, $-.036$] \\
Benefits $\rightarrow$ intention, total &
.440 & [.276, .600] & .451 & [.351, .552] \\
$R^2$, curricular devaluation &
.476 & --- & .539 & --- \\
$R^2$, integrative intention &
.406 & --- & .424 & --- \\
\hline
\end{tabular}
\vspace{0.5em}
\begin{minipage}{0.96\textwidth}
\small
\textit{Note.} MLR intervals for direct pathways are robust sandwich intervals; MLR intervals for indirect and total associations are percentile bootstrap intervals from 5,000 respondent-level resamples. WLSMV intervals are delta-method intervals because nonparametric resampling is not well defined for the polychoric input used by that estimator. The procedures are not equivalent, so intervals should be compared for direction and overlap rather than width. ns = interval includes zero.
\end{minipage}
\end{table}

\section{Discussion}
This study examined how perceived academic burden and perceived benefits were associated with students’ valuation of required broader coursework and their intentions to integrate that learning. The findings supported the proposed curricular legitimacy framework in three respects. Burden and benefits were associated with curricular devaluation in opposite directions, devaluation was associated with weaker integrative intention, and burden and benefits showed different combinations of direct and indirect associations with intention. The resulting pattern provides a more discriminating explanation than a workload-only account of student resistance.

\subsection{Interpreting the Asymmetric Appraisal Pathways}
The central result is not simply that burden was unfavourable and benefit was favourable. That conclusion would add little theoretically. The more informative finding is that burden and benefits occupied different positions in the estimated pathway structure.
Academic burden was strongly associated with greater curricular devaluation, and devaluation was associated with weaker integrative intention. Once this evaluative pathway was represented, the burden to intention pathway was not required to reproduce the observed covariance structure under maximum likelihood estimation, whereas the benefits pathway was, and the total association for burden was uncertain. Benefits, by contrast, were associated with less devaluation and retained an additional positive association with intention. The observed pattern is therefore consistent with burden being related to intention chiefly through how students evaluated the curricular domain, while benefits were related to intention through both weaker devaluation and an additional positive route.
This distinction is consistent with situated expectancy value theory, which treats perceived cost and task value as related but conceptually distinct appraisals situated within identities, goals, and institutional contexts \cite{eccles2020expectancy, Flake2015}. Effort is not inherently delegitimizing. Students can regard demanding work as worthwhile when its contribution to valued goals is credible. Conversely, even a manageable course can be devalued when its professional purpose remains unclear. Students’ workload judgements are therefore shaped not only by the quantity of work but also by how its purpose, organisation, and relevance are interpreted \cite{Kember2004}.
The findings should not be read as proof of a causal psychological mechanism. The cross-sectional design cannot establish that burden preceded devaluation or that devaluation subsequently changed intention. The estimates show that the proposed indirect configuration is consistent with the observed covariance structure and retained its direction across the reported sensitivity analyses. The residual burden to intention pathway was estimator dependent and should not be used to claim that an additional direct association is absent.

\subsection{Curricular Legitimacy as the Evaluative Bridge}

The concept of a curricular legitimacy gap identifies a problem that additive curriculum reform can overlook. Requiring communication, ethics, humanities, social science, or organisational coursework establishes curricular presence, but it does not ensure that students regard the associated knowledge as part of computing expertise.
This distinction matters in technical programmes where disciplinary cultures can separate legitimate engineering knowledge from social, ethical, and public considerations \cite{Cech2014, Lim2021, Niles2020}. When broader learning appears only in an isolated service course and does not reappear in programming assignments, system design, artificial intelligence projects, cybersecurity cases, internships, or capstones, the enacted curriculum may contradict the formal curriculum. The programme formally requires the content while implicitly signalling that it is dispensable in consequential technical work.
The negative association between curricular devaluation and integrative intention isolates the evaluative judgement connecting students’ curricular experience with their anticipated use of the learning. This finding suggests that reluctance cannot be understood only as dislike, difficulty, or overload. The relevant question is what students take the burden to signify. Reducing assessment volume may address avoidable pressure, but it will not necessarily establish professional legitimacy. A lighter course can remain peripheral, while demanding learning can be valued when students repeatedly observe its consequences for technical judgement.
Evidence from humanities and engineering interventions similarly indicates that students’ sociotechnical reasoning can be strengthened when human and contextual considerations are integrated into substantive educational activity rather than merely positioned beside technical content \cite{Czerwionka2025}. This interpretation is also consistent with scholarship that treats interdisciplinarity as the integration of perspectives and methods, not merely the curricular coexistence of different subjects \cite{Ming2026, Spelt2009}.
This programme-level interpretation is consistent with Embedded EthiCS, which places ethical reasoning within core computer science courses instead of assigning it exclusively to a stand-alone requirement \cite{Grosz2019}. The present model adds a testable explanation for why recurrence may matter: repeated use in consequential technical work can signal that sociotechnical reasoning belongs within computing expertise and can make its utility observable.

\subsection{Theoretical Contribution and Novelty}
The paper does not claim to be the first study of computing students’ responses to ethics, professional skills, or broader learning. Recent reviews already document diverse forms of student acceptance and resistance and the uneven evaluation of computing ethics instruction \cite{brown2024teaching, Padiyath2024}. The contribution is a reformulation of the explanandum: not whether a particular intervention is liked, but whether an institutionally required domain is accorded professional legitimacy and how that valuation is associated with intended reuse.
The resulting compliance, valuation, intention, and behaviour sequence provides the first contribution. It prevents compulsory course completion from being treated as educational acceptance and stated intention from being presented as observed transfer. The second contribution is construct precision. Curricular devaluation describes ambiguity, perceived waste, lack of necessity, and relative irrelevance, whereas academic resistance would imply protest, avoidance, affective opposition, or behaviour that the instrument did not measure.
The third contribution is the asymmetric appraisal architecture. Under the primary specification, burden was associated strongly with devaluation but had no additional supported association with intention, whereas benefits were associated both with lower devaluation and with stronger intention directly. The ordinal sensitivity model qualifies the first part of that contrast because the residual burden pathway became positive and significant. The robust conclusion is therefore not that burden has no direct pathway, but that burden and benefit do not behave as mirror image appraisals and that their indirect routes through curricular devaluation differ in sign.
The empirical contribution lies in evaluating this constructually refined, path-specific model in a technically intensive and internationally underrepresented CSE setting while reporting nonnormality, comparative measurement fit, measurement limitations, bootstrap instability, response quality checks, and estimator sensitivity. The evidence is best described as an initial theory test with transparent boundaries, not as definitive scale validation or causal mediation.

\subsection{Robustness and Evidential Boundaries}
The sensitivity analyses strengthened the evidential basis of the study without eliminating its boundaries. The observed two-item benefits index yielded complete convergence and admissible solutions across all 5,000 bootstrap resamples, whereas the two-indicator latent specification did not. Retaining the latent benefits factor as a specification check showed that the principal coefficient directions and hypothesis decisions did not depend on that modelling choice. Excluding invariant response vectors and respondents with dominant category use produced the same substantive pathway configuration. Ordinal estimation also preserved the directions and indirect associations, although the residual burden to intention pathway became positive and statistically distinguishable from zero. Claims about that direct pathway are therefore estimator dependent.
These checks support the stability of the observed pattern, but they do not convert the study-specific measures into mature scales. Reliability for perceived benefits and integrative intention was modest, AVE remained below .50 for three constructs, and robust TLI remained slightly below .90. The appropriate contribution is therefore an initial test of a theoretically specified model and a clarification of the measured constructs, not definitive scale validation or confirmation of a causal mechanism.

\section{Implications for Curriculum and Teaching}
The findings suggest a dual curriculum design strategy. Programmes should reduce avoidable ambiguity and coordination costs that may be interpreted as evidence of curricular irrelevance. At the same time, they should create direct and repeated experiences showing that broader learning has consequences for technical judgement. The objective is not to make broader coursework intellectually easier. It is to make its purpose, epistemic value, and professional consequences visible.

Table~\ref{tab:curriculum_design_propositions} translates the asymmetric appraisal model into evidence-informed curriculum design propositions that can be evaluated in future interventions.

\begin{table}[htbp]
\centering
\caption{Curriculum design propositions derived from the asymmetric appraisal model}
\label{tab:curriculum_design_propositions}
\renewcommand{\arraystretch}{1.2}
\setlength{\tabcolsep}{5pt}
\begin{tabular}{p{0.20\textwidth} p{0.38\textwidth} p{0.30\textwidth}}
\hline
\textbf{Design priority} &
\textbf{Concrete curriculum action} &
\textbf{Mechanism addressed} \\
\hline
Make professional utility consequential &
Require ethical, social, communicative, or organisational analysis to alter a system requirement, model choice, interface decision, deployment plan, or risk judgement. &
Makes benefits visible, may reduce devaluation, and provides an affirmative reason for later integration. \\
\hline
Reinforce legitimacy across the programme &
Reuse sociotechnical criteria in programming assignments, design reviews, artificial intelligence projects, cybersecurity cases, internships, and capstones. &
Signals that broader learning belongs within technical formation rather than within an isolated requirement. \\
\hline
Reduce ambiguity rather than intellectual demand &
Explain why each task belongs in the programme, coordinate deadlines, clarify assessment standards, and scaffold movement between quantitative and interpretive reasoning. &
May reduce the translation of avoidable academic burden into curricular devaluation. \\
\hline
Preserve disciplinary rigour &
Teach relevant humanities and social science methods using explicit evidence standards rather than presenting the content as generic soft skills. &
Strengthens epistemic credibility and reduces the appearance of superficial curricular inclusion. \\
\hline
Assess integration rather than attitude alone &
Use design rationales, stakeholder analyses, scenario-based assessments, portfolios, internship reflections, and later project artefacts alongside surveys. &
Tests whether integrative intention becomes observable behavioural integration. \\
\hline
Audit programme-level curricular signals &
Map where sociotechnical learning outcomes are introduced, revisited, and assessed across the whole programme, and identify the technical courses in which they are absent from assessment criteria and design decisions. &
Exposes divergence between the formal and the enacted curriculum, where later technical courses signal that broader learning is dispensable in consequential work. \\
\hline
\end{tabular}
\vspace{0.5em}
\begin{minipage}{0.94\textwidth}
\small
\textit{Note.} These propositions are derived from cross-sectional associations and should be evaluated through longitudinal or experimental curriculum research before being interpreted as established causal interventions.
\end{minipage}
\end{table}

For instructors, the practical test is whether broader learning changes what students do in a technical task. An ethics discussion that never affects a design decision is unlikely to establish legitimacy. By contrast, an assessment requiring students to identify stakeholders, challenge assumptions, justify tradeoffs, and revise a technical solution makes the professional consequence of broader learning observable.

Programme leaders should also examine the cumulative sequence of curricular signals. A stand-alone non-STEM course cannot carry the full responsibility for integration if later technical subjects treat its concepts as irrelevant. Professional utility should be demonstrated repeatedly through assessment, technical decision making, industry projects, internships, and capstone work.

The weak and estimator-dependent direct burden pathway also argues against treating workload reduction as the sole intervention. Coordination, clarity, and manageable assessment remain important, but a programme can reduce workload while leaving the legitimacy gap intact. Cost management and utility development should therefore be treated as complementary rather than interchangeable curriculum strategies.

\section{Limitations and Future Research}
\label{sec:limitations}
Several limitations define the boundaries of the study’s contribution. First, the cross-sectional design does not establish temporal precedence among academic appraisals, curricular devaluation, and integrative intention. Although the ordering of the structural pathways was derived theoretically, the direct and indirect estimates remain associational and should not be interpreted as evidence of causal mediation \cite{Maxwell2007}. Reverse and reciprocal relationships are also possible. For example, students who are already inclined to integrate broader learning may subsequently perceive greater benefits and express less curricular devaluation.
Second, the recruitment strategy limits statistical generalisation. Participants were recruited through non-probability snowball sampling, so the invitation propagated through students’ own networks rather than from a defined sampling frame. The number of students who received or viewed it cannot be determined, and neither a conventional response rate nor a direct assessment of nonresponse bias can therefore be reported; snowball propagation may also concentrate the sample within connected peer groups, compounding the self-selection inherent in voluntary participation. The sample was further concentrated among third- and fourth-year students, with first-year students excluded by design because they have typically had limited exposure to the broader coursework component of the degree. The findings consequently describe students who have already accumulated substantial experience of that coursework and provide no evidence about how curricular valuation first forms. Institutional affiliation was not collected at any stage, so the number of institutions represented, institution-level sample sizes, institutional clustering, and differences in programme design could not be examined.
Third, the measurement evidence was adequate for an initial test of the proposed model but does not support claims of definitive scale validation. All focal constructs were measured through self-reports collected within the same questionnaire, creating possible common source and consistency effects \cite{Podsakoff2003}. These effects may be compounded by the instrument itself: 12 of the 15 items displayed the label of their intended subconstruct to participants, which could encourage consistent responding within a labelled group and inflate within-construct correlations. The outcome represents integrative intention rather than observed educational or professional behaviour. Perceived benefits was measured using only two items, reliability for integrative intention was modest, and average variance extracted remained below .50 for three constructs. In addition, the participant-facing category “non-STEM” combined courses that may differ substantially in content, pedagogy, assessment, and proximity to technical practice. English-only administration without cognitive or linguistic pretesting may also have affected item interpretation and who chose to participate. The stability observed across the measurement and response-quality sensitivity analyses strengthens confidence in the structural pattern, but it does not remove these measurement constraints.
Future research should address these limitations through a staged programme of measurement development and prospective testing. The measures should first be refined through expert review, cognitive interviewing, expanded item pools, course-specific referents, and independent validation samples (Boateng et al., 2018). Subsequent longitudinal studies should assess appraisals, curricular devaluation, and integrative intention at multiple stages of the degree and connect them with observable evidence from design artefacts, stakeholder analyses, internships, and capstone projects. Finally, preregistered curriculum interventions could compare standard course framing with explicit professional utility framing and utility framing combined with repeated application in technical work. Such designs would provide a stronger test of whether visible utility reduces curricular devaluation, whether improved coordination weakens the association between burden and devaluation, and whether integrative intention develops into demonstrated professional practice.

\section{Conclusion}
Required coursework can appear on a transcript without acquiring a legitimate place in students’ conception of computing expertise. The findings indicate that academic burden and perceived benefits were associated with integrative intention through different pathway configurations. Burden was associated with stronger curricular devaluation, which was in turn associated with weaker intention. Benefits were associated with weaker devaluation and retained an additional positive association with intention. This asymmetric pattern shifts the explanatory focus beyond workload alone. It suggests that the educational significance of curricular cost depends partly on the meaning students assign to that cost, while visible professional utility provides both a basis for legitimacy and an affirmative reason to reuse the learning. The practical implication is not simply to reduce the difficulty of broader coursework, but to ensure that it repeatedly influences consequential technical decisions. The evidence remains cross-sectional, recruited by convenience and snowball sampling, and dependent on developing measures. Nevertheless, the curricular legitimacy framework provides a conceptually precise and empirically testable basis for longitudinal, experimental, and mixed methods research. Its central proposition is that curricular inclusion becomes educational integration only when broader forms of knowledge are recognised and enacted as legitimate components of computing practice.

\section*{Declarations}

\subsection*{Ethics approval and consent to participate}
Ethical approval was granted by the Research Ethics Committee of the Faculty of Science and Information Technology, Daffodil International University (reference IEC-FSIT/DIU/2025/2003; approved 2 April 2025) for the proposal ``STEM Students' Grumbling Attitude Towards Non-STEM Courses''. All participants provided informed electronic consent before entering the questionnaire, and a separate optional consent item covered publication and verbatim quotation of open-response comments. Participation was restricted to students aged 18 years or older.

\subsection*{Data availability}
The anonymous dataset analysed in this study is openly available in Mendeley Data under a Creative Commons Attribution 4.0 licence, version 5, \url{https://doi.org/10.17632/phk8yrfrc4.5}. The version-specific DOI is reported so that the analytical source remains reproducible if the repository record is updated.

\subsection*{Code availability}
The complete R analysis script, model syntax, package requirements, and random seed are supplied as supplementary code and will be archived with the full output in a persistent public repository. The archived output will include \texttt{sessionInfo()} documenting the exact R and package versions used for the final run; the analysis seed is 20260807.

\subsection*{Consent for publication}
Not applicable. The study collected no names, student identification numbers, email addresses, or IP addresses, and no individually identifiable data, images, or case details are reported. Participants who supplied open-response comments provided separate optional consent for those comments to be published in the open dataset and quoted verbatim in research publications; only comments covered by that consent are eligible for quotation.

\subsection*{Competing interests}
The authors declare that they have no competing financial or non-financial interests.

\subsection*{Funding}
This research received no specific grant from any funding agency in the public, commercial, or not-for-profit sectors.

\subsection*{Author contributions}
Walayat Hussain: conceptualization, methodology, validation, writing, review and editing, supervision. Md Abdullah Al Kafi: conceptualization, methodology, investigation, data curation, formal analysis, software, visualization, writing, original draft, project administration. Raka Moni: writing, original draft. All authors read and approved the final manuscript.

\subsection*{Declaration of generative AI and AI-assisted technologies in the manuscript preparation process}
Generative artificial intelligence tools, including ChatGPT (OpenAI) and Gemini (Google), were used for language refinement, manuscript organisation, and code review. All analyses, interpretations, citations, and final wording were reviewed and verified by the authors, who take full responsibility for the manuscript.

\bibliographystyle{plainnat}
\bibliography{references}

@article{Anderson1988,
   author = {James C. Anderson and David W. Gerbing},
   doi = {10.1037/0033-2909.103.3.411},
   issn = {1939-1455},
   issue = {3},
   journal = {Psychological Bulletin},
   month = {5},
   pages = {411-423},
   title = {Structural equation modeling in practice: A review and recommended two-step approach.},
   volume = {103},
   year = {1988}
}

@article{eccles2020expectancy,
  title={From expectancy-value theory to situated expectancy-value theory: A developmental, social cognitive, and sociocultural perspective on motivation},
  author={Eccles, Jacquelynne S and Wigfield, Allan},
  journal={Contemporary educational psychology},
  volume={61},
  pages={101859},
  year={2020},
  publisher={Elsevier}
}

@misc{Kline,
	title = {Principles and {Practice} of {Structural} {Equation} {Modeling}: {Fifth} {Edition}},
	shorttitle = {Principles and {Practice} of {Structural} {Equation} {Modeling}},
	url = {https://www.guilford.com/books/Principles-and-Practice-of-Structural-Equation-Modeling/Rex-Kline/9781462551910},
	language = {en-US},
	urldate = {2026-08-15},
	journal = {Guilford Press},
}

@article{meade_identifying_2012,
	address = {US},
	title = {Identifying careless responses in survey data},
	volume = {17},
	issn = {1939-1463},
	doi = {10.1037/a0028085},
	number = {3},
	journal = {Psychological Methods},
	publisher = {American Psychological Association},
	author = {Meade, Adam W. and Craig, S. Bartholomew},
	year = {2012},
	pages = {437--455},
}

@article{kopelman_model_1983,
	title = {A model of work, family, and interrole conflict: {A} construct validation study},
	volume = {32},
	issn = {0030-5073},
	shorttitle = {A model of work, family, and interrole conflict},
	url = {https://www.sciencedirect.com/science/article/pii/0030507383901472},
	doi = {10.1016/0030-5073(83)90147-2},
	number = {2},
	urldate = {2026-08-15},
	journal = {Organizational Behavior and Human Performance},
	author = {Kopelman, Richard E. and Greenhaus, Jeffrey H. and Connolly, Thomas F.},
	month = oct,
	year = {1983},
	pages = {198--215},
}

@article{brown2024teaching,
  title={Teaching ethics in computing: A systematic literature review of ACM computer science education publications},
  author={Brown, Noelle and Xie, Benjamin and Sarder, Ella and Fiesler, Casey and Wiese, Eliane S},
  journal={ACM Transactions on Computing Education},
  volume={24},
  number={1},
  pages={1--36},
  year={2024},
  publisher={ACM New York, NY}
}

@article{Bedewy2015,
   author = {Dalia Bedewy and Adel Gabriel},
   doi = {10.1177/2055102915596714},
   issn = {2055-1029},
   issue = {2},
   journal = {Health Psychology Open},
   month = {7},
   title = {Examining perceptions of academic stress and its sources among university students: The Perception of Academic Stress Scale},
   volume = {2},
   year = {2015}
}

@article{Boateng2018,
   author = {Godfred O. Boateng and Torsten B. Neilands and Edward A. Frongillo and Hugo R. Melgar-Quiñonez and Sera L. Young},
   doi = {10.3389/fpubh.2018.00149},
   issn = {2296-2565},
   journal = {Frontiers in Public Health},
   month = {6},
   title = {Best Practices for Developing and Validating Scales for Health, Social, and Behavioral Research: A Primer},
   volume = {6},
   year = {2018}
}

@article{Byrne2020,
   author = {Zinta S. Byrne and James W. Weston and Kelly Cave},
   doi = {10.1007/s11165-018-9738-3},
   issn = {0157-244X},
   issue = {4},
   journal = {Research in Science Education},
   month = {8},
   pages = {1417-1433},
   title = {Development of a Scale for Measuring Students’ Attitudes Towards Learning Professional (i.e., Soft) Skills},
   volume = {50},
   year = {2020}
}

@article{Cech2014,
   author = {Erin A. Cech},
   doi = {10.1177/0162243913504305},
   issn = {0162-2439},
   issue = {1},
   journal = {Science, Technology, \& Human Values},
   month = {1},
   pages = {42-72},
   title = {Culture of Disengagement in Engineering Education?},
   volume = {39},
   year = {2014}
}

@article{Czerwionka2025,
   author = {Lori Czerwionka and Siddhant S. Joshi and Gabriel O. Rios‐Rojas and Kirsten A. Davis},
   doi = {10.1002/jee.70004},
   issn = {1069-4730},
   issue = {3},
   journal = {Journal of Engineering Education},
   month = {7},
   title = {Developing students' sociotechnical thinking in the humanities‐engineering classroom: A treatment versus control study using a scenario‐based assessment},
   volume = {114},
   year = {2025}
}

@article{Dennis2010,
   author = {John P. Dennis and Jillon S. Vander Wal},
   doi = {10.1007/s10608-009-9276-4},
   issn = {0147-5916},
   issue = {3},
   journal = {Cognitive Therapy and Research},
   month = {6},
   pages = {241-253},
   title = {The Cognitive Flexibility Inventory: Instrument Development and Estimates of Reliability and Validity},
   volume = {34},
   year = {2010}
}

@article{Flake2015,
   author = {Jessica Kay Flake and Kenneth E. Barron and Christopher Hulleman and Betsy D. McCoach and Megan E. Welsh},
   doi = {10.1016/j.cedpsych.2015.03.002},
   issn = {0361476X},
   journal = {Contemporary Educational Psychology},
   month = {4},
   pages = {232-244},
   title = {Measuring cost: The forgotten component of expectancy-value theory},
   volume = {41},
   year = {2015}
}

@article{Grosz2019,
   author = {Barbara J. Grosz and David Gray Grant and Kate Vredenburgh and Jeff Behrends and Lily Hu and Alison Simmons and Jim Waldo},
   doi = {10.1145/3330794},
   issn = {0001-0782},
   issue = {8},
   journal = {Communications of the ACM},
   month = {7},
   pages = {54-61},
   title = {Embedded EthiCS},
   volume = {62},
   year = {2019}
}

@incollection{Hart,
	series = {Advances in {Psychology}},
	title = {Development of {NASA}-{TLX} ({Task} {Load} {Index}): {Results} of {Empirical} and {Theoretical} {Research}},
	volume = {52},
	shorttitle = {Development of {NASA}-{TLX} ({Task} {Load} {Index})},
	url = {https://www.sciencedirect.com/science/article/pii/S0166411508623869},
	doi = {10.1016/S0166-4115(08)62386-9},
	urldate = {2026-08-16},
	booktitle = {Human {Mental} {Workload}},
	publisher = {North-Holland},
	author = {Hart, Sandra G. and Staveland, Lowell E.},
	editor = {Hancock, Peter A. and Meshkati, Najmedin},
	month = jan,
	year = {1988},
	pages = {139--183},
}

@article{Henseler2015,
   author = {Jörg Henseler and Christian M. Ringle and Marko Sarstedt},
   doi = {10.1007/s11747-014-0403-8},
   issn = {0092-0703},
   issue = {1},
   journal = {Journal of the Academy of Marketing Science},
   month = {1},
   pages = {115-135},
   title = {A new criterion for assessing discriminant validity in variance-based structural equation modeling},
   volume = {43},
   year = {2015}
}

@article{Kember2004,
   author = {David Kember *},
   doi = {10.1080/0307507042000190778},
   issn = {0307-5079},
   issue = {2},
   journal = {Studies in Higher Education},
   month = {4},
   pages = {165-184},
   title = {Interpreting student workload and the factors which shape students' perceptions of their workload},
   volume = {29},
   year = {2004}
}

@book{Kumar2024,
   author = {Amruth N. Kumar and Rajendra K. Raj and Sherif G. Aly and Monica D. Anderson and Brett A. Becker and Richard L. Blumenthal and Eric Eaton and Susan L. Epstein and Michael Goldweber and Pankaj Jalote and Douglas Lea and Michael Oudshoorn and Marcelo Pias and Susan Reiser and Christian Servin and Rahul Simha and Titus Winters and Qiao Xiang},
   city = {New York, NY, USA},
   doi = {10.1145/3664191},
   isbn = {9798400710339},
   month = {1},
   publisher = {ACM},
   title = {Computer Science Curricula 2023},
   year = {2024}
}

@article{Li2016,
   author = {Cheng-Hsien Li},
   doi = {10.3758/s13428-015-0619-7},
   issn = {1554-3528},
   issue = {3},
   journal = {Behavior Research Methods},
   month = {9},
   pages = {936-949},
   title = {Confirmatory factor analysis with ordinal data: Comparing robust maximum likelihood and diagonally weighted least squares},
   volume = {48},
   year = {2016}
}

@article{Lim2021,
   author = {Jae Hoon Lim and Brittany D. Hunt and Nickcoy Findlater and Peter T. Tkacik and Jerry L. Dahlberg},
   doi = {10.1007/s11948-021-00355-0},
   issn = {1353-3452},
   issue = {6},
   journal = {Science and Engineering Ethics},
   month = {12},
   pages = {74},
   title = {“In Our Own Little World”: Invisibility of the Social and Ethical Dimension of Engineering Among Undergraduate Students},
   volume = {27},
   year = {2021}
}

@article{MARDIA1970,
   author = {K. V. MARDIA},
   doi = {10.1093/biomet/57.3.519},
   issn = {0006-3444},
   issue = {3},
   journal = {Biometrika},
   pages = {519-530},
   title = {Measures of multivariate skewness and kurtosis with applications},
   volume = {57},
   year = {1970}
}

@article{Maxwell2007,
   author = {Scott E. Maxwell and David A. Cole},
   doi = {10.1037/1082-989X.12.1.23},
   issn = {1939-1463},
   issue = {1},
   journal = {Psychological Methods},
   pages = {23-44},
   title = {Bias in cross-sectional analyses of longitudinal mediation.},
   volume = {12},
   year = {2007}
}

@article{Ming2026,
   author = {Xin Ming and Miles MacLeod and Mieke Boon and Jan van der Veen},
   doi = {10.1080/03043797.2025.2534646},
   issn = {0304-3797},
   issue = {2},
   journal = {European Journal of Engineering Education},
   month = {3},
   pages = {220-240},
   title = {Envisioning and interpreting interdisciplinary engineering education (IEE): different dimensions of understanding interdisciplinarity},
   volume = {51},
   year = {2026}
}

@article{Niles2020,
   author = {Skye Niles and Santina Contreras and Shawhin Roudbari and Jessica Kaminsky and Jill Lindsey Harrison},
   doi = {10.1002/jee.20323},
   issn = {1069-4730},
   issue = {3},
   journal = {Journal of Engineering Education},
   month = {7},
   pages = {491-507},
   title = {Resisting and assisting engagement with public welfare in engineering education},
   volume = {109},
   year = {2020}
}

@article{Padiyath2024,
   author = {Aadarsh Padiyath},
   doi = {10.1145/3639572},
   issn = {1946-6226},
   issue = {2},
   journal = {ACM Transactions on Computing Education},
   month = {6},
   pages = {1-19},
   title = {A Realist Review of Undergraduate Student Attitudes towards Ethical Interventions in Technical Computing Courses},
   volume = {24},
   year = {2024}
}

@article{Podsakoff2003,
   author = {Philip M. Podsakoff and Scott B. MacKenzie and Jeong-Yeon Lee and Nathan P. Podsakoff},
   doi = {10.1037/0021-9010.88.5.879},
   issn = {1939-1854},
   issue = {5},
   journal = {Journal of Applied Psychology},
   pages = {879-903},
   title = {Common method biases in behavioral research: A critical review of the literature and recommended remedies.},
   volume = {88},
   year = {2003}
}

@article{Rizzo1970,
   author = {John R. Rizzo and Robert J. House and Sidney I. Lirtzman},
   doi = {10.2307/2391486},
   issn = {00018392},
   issue = {2},
   journal = {Administrative Science Quarterly},
   month = {6},
   pages = {150},
   title = {Role Conflict and Ambiguity in Complex Organizations},
   volume = {15},
   year = {1970}
}

@article{Rosseel2012,
   author = {Yves Rosseel},
   doi = {10.18637/jss.v048.i02},
   issn = {1548-7660},
   issue = {2},
   journal = {Journal of Statistical Software},
   title = {lavaan : An R Package for Structural Equation Modeling},
   volume = {48},
   year = {2012}
}

@article{Schijf2023,
   author = {Jennifer E. Schijf and Greetje P.C. van der Werf and Ellen P.W.A. Jansen},
   doi = {10.1080/21568235.2022.2058045},
   issn = {2156-8235},
   issue = {4},
   journal = {European Journal of Higher Education},
   month = {10},
   pages = {429-447},
   title = {Measuring interdisciplinary understanding in higher education},
   volume = {13},
   year = {2023}
}

@article{Spelt2009,
   author = {Elisabeth J. H. Spelt and Harm J. A. Biemans and Hilde Tobi and Pieternel A. Luning and Martin Mulder},
   doi = {10.1007/s10648-009-9113-z},
   issn = {1040-726X},
   issue = {4},
   journal = {Educational Psychology Review},
   month = {12},
   pages = {365-378},
   title = {Teaching and Learning in Interdisciplinary Higher Education: A Systematic Review},
   volume = {21},
   year = {2009}
}

\end{document}